\documentstyle[epsf,12pt]{article}
\renewcommand{\cite}[1]{\ref{#1}}
\newcommand{\BBox}{\mbox{\raisebox{-0.2em}{\large$\Box$}}}
\newcommand{\half}{\frac{1}{2}}

\newcommand{\Dslash}{D\hspace{-.65em}/}

\newcommand{\ksl}{k\hspace{-.5em}/}
\newcommand{\psls}{p\hspace{-.4em}/}
\newcommand{\beq}{\begin{equation}}
\newcommand{\eeq}{\end{equation}}
\newcommand{\beqa}{\begin{eqnarray}}
\newcommand{\eeqa}{\end{eqnarray}}
\newcommand{\bpc}{\begin{picture}}
\newcommand{\epc}{\end{picture}}
\newcommand{\bfg}{\begin{figure}}
\newcommand{\efg}{\end{figure}}
\newcommand{\psibar}{\overline{\psi}}
\newcommand{\reflef}{(\ref}
\newcommand{\nnb}{\nonumber}

\begin{document}

\begin{center}
{\large\bf Role of Regularization in Quantum Corrections to the\vspace{.2em}
 Scalar-Tensor Theories of Gravity}\vspace{.3cm}\\
Yasunori Fujii\footnote{Electronic address: fujii@handy.n-fukushi.ac.jp}\\
Nihon Fukushi University, Handa, 475\ Japan\\
and\\
ICRR, University of Tokyo, Tanashi, Tokyo, 188 Japan \vspace{.5cm}\\
\end{center}
\bigskip
\begin{center}
{\large\bf Abstract}
\end{center}
\baselineskip=0.4cm
\hspace*{1cm}
\parbox{12.5cm}
{We show that regularizing divergent integrals is crucially important when applied to the loop diagrams corresponding to quantum corrections to the coupling of the ``gravitational" scalar field due to the interaction among matter fields.  We use the method of continuous spacetime dimensions to demonstrate that WEP is a robust property of the Brans-Dicke theory beyond the classical level, hence correcting our previous assertion of the contrary.  The same technique can be used to yield the violation of WEP when applied to the scale-invariant theory, thus providing another reason for expecting fifth-force-type phenomena.}
\vspace{2em}\\
\baselineskip=0.6cm

In our previous paper [\cite{yf1}] we argued that a type of quantum correction to the Brans-Dicke (BD) theory [\cite{bd}] results in a rather large effect of violating composition-independence of the scalar-matter coupling.  It was claimed that precision tests of WEP constrain the parameter $\omega$ much more strongly than those derived from the solar-system experiments.

Subsequently, however, we came to realize that the effect is an artifact of our neglect of ``regularizing" basically divergent integrals, thus re-establishing WEP as a robust property of the theory beyond the classical level.  The revised analysis to be presented here is essentially revisiting the lesson which was learned in such historic issues as self-stress [\cite{sstr}] and trace anomaly [\cite{tran}].   It seems, however, that the same care might be crucially important in studying other quantum effects to some of the matter-gravity interaction.  For this reason we try to explain essential ingredients of the discussion briefly but in a somewhat self-contained manner, hence allowing minimum overlapping with Ref. [\cite{yf1}].

According to the ``lesson," the fact that an integral which is divergent judged from power-counting turns out to be finite does not necessarily mean that no regularization procedure is required [\cite{tkhs}].  We use the method of continuous spacetime dimensions which is particularly convenient when gauge invariance is to be observed.

We start with the Lagrangian\footnote{$\xi$ is related to $\omega$ in the original notation [\cite{bd}] by $\xi =1/(4\omega)$.  We set $8\pi G_{0}=c=\hbar =1$, where $G_{0}=((6\xi +1)/(8\xi +1))G$, with $G$ the gravitational constant measured by the Cavendish-type experiment, including the scalar contribution.}
\beq
{\cal L}=\sqrt{-g}\left[ \half\xi\phi^2 R -\half g^{\mu\nu}
\partial_{\mu}\phi \partial_{\nu}\phi +L_{\rm matter}
\right],
\label{cr1-1}
\eeq
in $N$-dimensional spacetime.  For the sake of illustration we consider an example of the interacting system of electron and photon for the matter; 
\beq
{\cal L}_{\rm matter}={\cal L}_{\psi}+{\cal L}_{A},
\label{cr1-2}
\eeq
with
\beqa
{\cal L}_{\psi}&=&
-\sqrt{-g}\:\psibar\left(\mbox{\small$\frac{1}{2}$}( \Dslash-\stackrel{\leftarrow}{\Dslash} )+ m \right)\psi, \label{cr1-3}\\
{\cal L}_{A}&=& -\half \sqrt{-g}\:g^{\mu\nu}g^{\rho\sigma}
\partial_{\mu}A_{\rho}\partial_{\nu}A_{\sigma}+\cdots,\label{cr1-4}
\eeqa
where
\beq
D_{\mu}=\partial_{\mu}+ieA_{\mu}+\cdots,
\label{cr1-5}
\eeq
while we use the Feynman gauge, ignoring gravitationally higher-order terms.

In order that the standard technique of quantum electrodynamics can be applied we move to a conformal frame (Einstein frame) in which no nonminimal coupling is present.  For a conformal transformation with
\beq
g_{\mu\nu}=\Omega^{-2}g_{*\mu\nu}, 
\label{cr1-5a}
\eeq
where 
\beq
\Omega =\left( \xi\phi^2 \right)^{1/(N-2)},
\label{cr1-5b}
\eeq
we find
\beq
R=\Omega^2 \left[
R_{*}+ 2(N-1)\BBox_{*}f -(N-1)(N-2)
(\nabla_{*\mu}f)(\nabla_{*}^{\mu}f)\right],
\label{cr1-6}
\eeq
where
\beq
f=\ln \Omega.
\label{cr1-7}
\eeq

The scalar field $\sigma$ which is canonical in E-frame is related to the original $\phi$ by
\beq
\phi =\xi^{-1/2}e^{\beta\sigma},
\label{cr1-8}
\eeq
where\footnote{$\sigma$ can be a normal field (with positive energy) even if $\phi$ is a ghost in the original frame, as with some model of Kaluza-Klein theory. See Ref. [\cite{yf2}]}
\beq
\beta^{-1}=\sqrt{4\frac{N-1}{N-2}+\xi^{-1}}.
\label{cr1-9}
\eeq
Notice that $\beta$ is in fact multiplied by $\sqrt{8\pi G_{0}}$ in the usual notation.  From \reflef{cr1-5b}) and \reflef{cr1-8}) we also find
\beq
\Omega =\exp\left( \frac{1}{\nu -1}\beta\sigma \right),
\label{cr1-10}
\eeq
with $\nu \equiv N/2$.

The kinetic part of $\psi$ can be made canonical by introducing $\psi_{*}$ by
\beq
\psi = \Omega^{(N-1)/2}\psi_{*},\quad {\rm and} \quad
\psibar = \Omega^{(N-1)/2}\psibar_{*}.
\label{cr1-11}
\eeq
The mass term is then transformed to
\beq
-me^{-\beta\sigma/(\nu -1)}\overline{\psi}_{*}\psi_{*}.
\label{cr1-12}
\eeq

On the other hand, ${\cal L}_{A}$, which is conformally invariant as it stands in 4 dimensions, is made also canonical in terms of $A_{*\mu}$ defined by
\beq
A_{\mu}=\Omega^{\nu-2}A_{*\mu}.
\label{cr1-13}
\eeq
Substituting this into \reflef{cr1-4}) will yield terms with $\sim \partial_{\mu}\sigma$, which can be ignored when we try to determine the matter coupling of $\sigma$ that carries almost zero 4-momentum, hence leaving the resulting equations nearly as simple as with the original $A_{\mu}$.  Using \reflef{cr1-13}) in \reflef{cr1-5}) gives, however,
\beq
D_{\mu}=\partial_{\mu}+ie e^{(\nu-2)\beta\sigma}A_{*\mu}+\cdots,
\label{cr1-13a}
\eeq
indicating the emergence of new terms of the $\sigma$ coupling; $(\nu -2)^{n}\psibar_{*} \gamma^{\mu}\psi_{*} A_{*\mu}\sigma^n$ with $n=1,2,3,\cdots$.  In practice we are concerned with $n=1$ only.  In this sense we follow two approaches either computing the trace of the energy-momentum tensor or evaluating the linear $\sigma$ couplings directly.

We first obtain the energy-momentum tensor in E-frame:
\beqa
T_{*\mu\nu}&=& -\frac{2}{\sqrt{-g_{*}}}\frac{\partial {\cal L}_{*\rm m}}{\partial g^{\mu\nu}_{*}}\nonumber\\
&=&\half \overline{\psi}_{*}\left(
 \gamma_{*(\mu}D_{*\nu )}-\stackrel{\leftarrow}{D}_{*(\mu}\gamma_{*\nu )}
\right)\psi_{*} \nonumber\\
 &&+g^{\rho\sigma}_{*}\left(
\partial_{\mu}A_{*\rho}\partial_{\nu}A_{*\sigma}\!
+\!\partial_{\rho}A_{*\mu}\partial_{\sigma}A_{*\nu}\right)
-\half g_{*\mu\nu} g^{\kappa\lambda}_{*}g^{\rho\sigma}_{*}
\partial_{\kappa}A_{*\rho}\partial_{\lambda}A_{*\sigma},
\label{cr1-16}
\eeqa
hence giving
\beq
-T_{*}= me^{-\beta\sigma/(\nu -1)}
\overline{\psi}_{*}\psi_{*} +(\nu -2)g^{\mu\nu}_{*}g^{\rho\sigma}_{*}\partial_{\mu}A_{*\rho}\partial_{\nu}A_{*\sigma}.
\label{cr1-17}
\eeq

In the original conformal frame we have the field equation for $\varphi\equiv (\xi/2)\phi^2$:\footnote{This $\varphi$ is in fact $\phi$ in the original notation in [\cite{bd}].}
\beq
\BBox\varphi =\beta^2 T,
\label{cr1-18}
\eeq
with $T$ the trace of the matter energy-momentum tensor.  Based on \reflef{cr1-16}) and the corresponding definition in the original conformal frame we derive
\beq
T_{\mu\nu}=\Omega^2 T_{*\mu\nu},
\label{cr1-18a}
\eeq
as long as ${\cal L}_{\rm matter}$ contains no derivatives of the metric, which condition is not obeyed in general but is justified in calculating the trace part.  On this basis we re-express \reflef{cr1-18}) in the form;
\beq
\BBox_{*}\sigma =\beta T_{*}.
\label{cr1-19}
\eeq
Using these equations we are able to verify the complete equivalence between the calculations in both conformal frames as far as we confine ourselves to the first-order expansion with respect to $\sqrt{G_{0}}\sigma$, though possibly entirely different physics may be expected in general.

We now compute the matrix elements of $T_{*}$ between $\psibar_{*}$ and $\psi_{*}$, and between $A_{*\mu}$ and $A_{*\nu}$, also to the first-order approximation in $\sqrt{G_{0}}$.  We are concerned also with the limit of zero 4-momentum transferred to $\sigma$.  In what follows we suppress the symbol $*$ to simplify the notation.

At the tree-level in 4 dimensions we have an obvious result:
\beq
-T_{0}=m\psibar\psi.
\label{cr1-20}
\eeq
The fact that this is the trace of the energy-momentum tensor should be extended to any complex system, hence guranteeing composition-independence of the scalar force.

To the next 1-loop approximation we consider the diagrams (a's) and (b's) in Fig. \cite{fg1}, arising from the first term (with $e^{-\beta\sigma/(\nu -1)}$ dropped) and the second terms of \reflef{cr1-17}), respectively.

Notice that $\nu-2$ in \reflef{cr1-17}) is kept nonzero until all the calculations are completed according to the recipe of dimensional regularization.  To analyze these diagrams it is most convenient to work with the self-energy diagrams (c1) and (c2) in Fig. \cite{fg2} for $\psi$ and $A_{\mu}$, respectively, and the corresponding functions:
\beqa
\Sigma(m,i\psls)&=&-i\frac{\alpha}{4\pi^{3}}\int d^{N}k
\gamma_{\mu}\frac{1}{\left( m+i\ksl  \right)(k-p)^{2}}\gamma^{\mu},\label{cr1-21}\\
\nnb\\
\Pi_{\mu\nu}(k)&=&-i\frac{\alpha}{4\pi^{3}}\int d^{N}p{\rm Tr}
\left[ \gamma_{\mu}\frac{1}{ m+i\psls}\gamma_{\nu}
\frac{1}{m+i\psls+i\ksl}\right].
\label{cr1-22}
\eeqa

As usual \reflef{cr1-21}) is  also expanded as
\beq
\Sigma(m,i\psls)=A(m)+\left( m+i\psls \right)B(m)
+\left( m+i\psls \right)^{2}\Sigma_{f}(m,i\psls),
\label{cr1-23}
\eeq
while gauge invariance in \reflef{cr1-22}) is manifest in the form
\beq
\Pi_{\mu\nu}(k)=C(m^{2},k^{2})\left( k_{\mu}k_{\nu}-k^{2}\eta_{\mu\nu} \right).
\label{cr1-24}
\eeq

The diagram (a1) represents a mass insertion to the diagram (c1), hence giving 
\beq
-<T>_{\psi}^{(a)}=\left( m\frac{\partial}{\partial m} \Sigma(m,i\psls)\right)_{i\psls =-m}.
\label{cr1-25}
\eeq
We include the contributions from diagrams (a$1'$) and (a$1''$).  Strictly speaking we must have added the left-right symmetric diagrams, but the final result should remain the same according to the spirit of wave-function renormalization.  In this way we arrive at a simple result:
\beq
-<T>_{\psi}^{(a)}= m\frac{dA(m)}{dm}.
\label{cr1-26}
\eeq

According to the standard calculation $A$ is given by
\beq
A=-i\frac{\alpha}{4\pi^3}m\int dx \int d^{N}k
\frac{N+x(2-N)}{(k^2 +a^2)^2},
\label{cr1-27}
\eeq
where
\beq
a^2=(1-x)^2 m^2.
\label{cr1-28}
\eeq
Using the integration formula
\beq
\int d^{N} k 
\frac{ ( k^2 )^{m-2}}{ (k^2 +a^2)^{n}}
= i\half \frac{2 \pi^{\nu}}{\Gamma (\nu)}
 \left( a^2 \right)^{\nu -n+m-2}
\frac{\Gamma(m-2+\nu)\Gamma(n-m+2-\nu)}{\Gamma(n)},
\label{cr1-29}
\eeq
we obtain
\beq
A=\frac{\alpha}{4\pi}m\Gamma(2-\nu)
\int dx (a^2)^{\nu -2}\left( N+x(2-N) \right).
\label{cr1-31}
\eeq
We substitute this into \reflef{cr1-26}) obtaining
\beqa
-<T>_{\psi}^{(a)}&=&A+\frac{\alpha}{4\pi}m\Gamma(2-\nu)
\int dx (\nu -2)m\left( a^2 \right)^{\nu -3}
\frac{d a^2}{dm}\left( N+x(2-N) \right)
\nonumber\\
&=&A-\frac{\alpha}{4\pi}2m^3 (2-\nu)\Gamma(2-\nu)
\int dx (1-x)^2 \left( a^2\right)^{\nu-3}
\left( N+x(2-N) \right)\nonumber\\
&\approx&A-\frac{\alpha}{4\pi}\Gamma(3-\nu)2m
\int dx \left( 4-2x   \right)\nonumber\\
&=&A-\frac{3\alpha}{2\pi}m,
\label{cr1-32}
\eeqa
which is the same result as in Ref. [\cite{yf1}].  Notice that the factor $(2-\nu)$ in \reflef{cr1-34}) cancels the pole of $\Gamma(2-\nu)$ thus giving a finite result;
\beq
(2-\nu)\Gamma(2-\nu)=\Gamma(3-\nu)\rightarrow \Gamma(1)=1.
\label{cr1-35}
\eeq

In fact $A=\delta m$ in the first term in the last line of \reflef{cr1-32}) gives the correction to the mass $m$ in the zeroth-order term \reflef{cr1-20}), whereas the second term $-(3\alpha /2\pi)m$, which is {\em not included} in $T$, violates composition-indepencence.

Now turn to the diagram (b1) in which we insert the second term of \reflef{cr1-17}).  We find $2(\nu -2)q^2$ with $q=p+k$ for the vertex while the photon propagator gives $q^{-2}$.  Also with the overall negative sign, which was responsible for the expression in \reflef{cr1-25}), we obtain
\beq
-<T>_{\psi}^{(b)}=-2(\nu -2)A,
\label{cr1-33}
\eeq 
with the same function $A$ as in \reflef{cr1-23}).  From \reflef{cr1-31}) we calculate as $N\rightarrow 4$;
\beqa
-<T>_{\psi}^{(b)}&=&\frac{\alpha }{2\pi}m (2-\nu)\Gamma(2-\nu)
\int dx (4-2x)\nnb\\
&=& \frac{3\alpha}{2\pi}m,
\label{cr1-34}
\eeqa
which exactly cancels the second term of \reflef{cr1-32}).

In this way we now have 
\beq
-<T>_{\psi}=m+\delta m =m_{\rm obs},
\label{cr1-36}
\eeq
now {\em respecting} composition-independence, up to the 1-loop order.

The same analysis is applied also to the photon matrix element $<T>_{\mu\nu}$ between the two states of the polarization $\mu$ and $\nu$.  The diagram (a2) (with the understanding that another diagram with the insertion at the other line is added) gives
\beq
-<T>_{\mu\nu}^{(a)}=m\frac{\partial}{\partial m}\Pi_{\mu\nu}(k),
\label{cr1-37}
\eeq
where we use \reflef{cr1-24}) with
\beq
C(k^2,m^2)=\frac{\alpha}{\pi}\Gamma(2-\nu)
\int dx \left( b^2 \right)^{\nu -2}2x(1-x),
\label{cr1-39}
\eeq
where
\beq
b^2 = x(1-x)k^2 +m^2.
\label{cr1-40}
\eeq

Using \reflef{cr1-24}) and \reflef{cr1-39}) in \reflef{cr1-37}) yields\beq
-<T>_{\mu\nu}^{(a)}=\left(  
k_{\mu}k_{\nu}-k^2 \eta_{\mu\nu}\right)\left(  m\frac{\partial C}{\partial m} \right)_{k^2 =0},
\label{cr1-41}
\eeq
where
\beqa
m\frac{\partial C}{\partial m}&=& 2m^2\frac{\partial C}{\partial m^2}\nnb\\
&=&2m^2\frac{\alpha}{\pi}\Gamma(2-\nu )
\int dx (\nu -2)\left( b^2 \right)^{\nu -3}
\frac{db^2}{dm^2} 2x(1-x) \nonumber\\
&=& 2m^2\frac{2\alpha}{\pi}(\nu -2)\Gamma(2-\nu )
\int dx \frac{x(1-x)}{m^2 +x(1-x)k^2}\nnb\\
&\stackrel{k^2 =0}{\longrightarrow}&-\frac{4\alpha}{\pi}
\Gamma(3-\nu)\int x(1-x)dx =-\frac{2\alpha}{3\pi},
\label{cr1-42}
\eeqa
which would indicate an effective term 
\beq
L'=-\frac{2\alpha}{3\pi}\beta \frac{1}{4}F_{\mu\nu}F^{\mu\nu}\sigma,
\label{cr1-43}
\eeq
in the Lagrangian, obviously a non-$T$ coupling of $\sigma$.

However, we now consider the diagram (b2) arising from the second term of \reflef{cr1-17}).  In the same way as in obtaining \reflef{cr1-26}) we find
\beq
-(\nu -2)2k^2\frac{1}{k^2}\Pi_{\mu\nu}(k)=\left( k_{\mu}k_{\nu}-k^2 \eta_{\mu\nu}\right) 2(2-\nu)C(k^2,m^2).
\label{cr1-44}
\eeq
As before again, $(2-\nu)$ cancels the pole at $\nu =2$ of $C$ as given by \reflef{cr1-39}), giving
\beq
2(2-\nu)C=\frac{\alpha}{3\pi}(2-\nu)\Gamma(2-\nu )=\frac{2\alpha}{3\pi},
\label{cr1-45}
\eeq
which cancels \reflef{cr1-42}), leaving no anomalous term like \reflef{cr1-43}).

Summarizing we conclude that composition-independence survives the correction up to the 1-loop order, suggesting the same to any order of perturbation expansion with respect to the interaction among matter fields.  The underlying reason is simple; the scalar field, which was chosen to be decoupled from the matter at the outset, couples finally to the matter via the ``mixing" (due to the nonminimal coupling) between the scalar field and the spin-0 part of the metric field that endorses composition-independence.

Mathematically, the contributions from the second term of \reflef{cr1-17}) are precisely what is known as the ``trace anomaly."\footnote{As one of the striking examples, we find that \reflef{cr1-45}) emerges hence giving nonzero $<T>_{\mu\nu}$ even if $m=0$.  This is justified if we let $\nu \rightarrow 2$ in \reflef{cr1-39}) and \reflef{cr1-40}) {\em before} setting $k^2= 0$.} [\cite{tran}]  This may sound a little confusing, however, because the same terms in our discussion act to remove the composition-dependent terms which might be also called ``anomaly" because they would break the property expected to hold classically.

On the other hand, the occurrence of ``self-stress," hence violating Lorentz covariance of an energy-momentum vector of the one-particle state, is related closely to the emergence of composition-dependence due to the first term of \reflef{cr1-17}); the extra term in \reflef{cr1-32}) has been simply identified with $T^{i}_{i}$ as part of $T^{\mu}_{\mu}$.  The same success in solving the problem by means of Pauli-Villars regulators [\cite{sstr},\cite{tkhs}] is enjoyed also by appealing to the method of continuous dimensions.

We reinforce our conclusion by evaluating the 1-loop correction to the linear $\sigma$ coupling without invoking the trace of the energy-momentum tensor explicitly.

We evaluate the same diagrams as (a1), (a1$'$),(a1$''$) and (a2).  Instead of (b1) and (b2), however, we consider the diagrams (d1) and (d2) in Fig. \cite{fg3}, which arise from the term in \reflef{cr1-13a}) linear in $\sigma$.  In effect we replace $e$ in the diagrams (c1) and (c2) in Fig. \cite{fg2} by
\beq
(\nu -2)\beta\sigma.
\label{cr1-46}
\eeq

We remove $\beta\sigma$ in accordance with the fact that the linear term in \reflef{cr1-12}) is $m\beta\sigma$, thus simply multiply $2(\nu -2)$ with $A$ and $C$, where the factor $2$ implies that we have the contributions from both vertices.

For the $\psi$ term, this is found to be the same as \reflef{cr1-34}), whereas the $A_{\mu}$ term gives the same term as \reflef{cr1-45}), reproducing the same result obtained from $T_{*}$.

Before closing the paper we add an interesting discussion on the same role of regularization in the ``scale-invariant theory" which is similar to but somewhat different from BD theory, serving as a basis of expecting a finite-range gravity due to a dilaton [\cite{yf5}].

Suppose we replace \reflef{cr1-3}) by
\beq
-\sqrt{-g}\:\psibar\left( \mbox{\small$\half$}( \Dslash -\stackrel{\leftarrow}{\Dslash})+ f\phi \right)\psi,
\label{cr1-47}
\eeq
where the coupling constant $f$ is dimensionless, hence providing a {\em scale-invariance} in the whole theory.  After moving to E frame, this interaction term is transformed to 
\beq
{\cal L}'=-\sqrt{-g_{*}}f\xi^{-1/2}\exp\left( \frac{\nu -2}{\nu-1}\beta\sigma \right)\psibar_{*}\psi_{*}.
\label{cr1-48}
\eeq
The first term of the expansion of the exponential gives a mass
\beq
m=f\xi^{-1/2},
\label{cr1-49}
\eeq
while other terms of $\sigma$ vanishes at $\nu =4$.  In this way we abandon one of the premises in BD theory by introducing the matter coupling of $\phi$ in the original conformal frame, but ending up with no matter coupling of $\sigma$ in E frame.  This might be an interesting trade-off in evading a serious consequence of BD theory in cosmology [\cite{yf2}].

As it turns out, however, the complete decoupling is purely classical.  The fact that decoupling occurs only at $\nu =2$ indicates that it may cease to hold once loop corrections are included giving poles at $\nu =2.$  This is shown to be the case, as will be sketched below.

For $\psi$ we consider the diagrams in Fig. 4.  In (e1) $\sigma$ couples through the second term of the expansion in \reflef{cr1-48}) whereas $\sigma$ comes in through the same term as in \reflef{cr1-13a}) in (e2) and (e3).
 Including diagrams of the type of wave-function renormalization is also understood in (e1) and (e2).  We find that $J_{1,2}$ as the source of $\sigma$ coming from (e1) and (e2) is given by
\beq
J_{1,2}=3\beta (\nu -2)m \frac{dA}{dm}.
\label{cr1-50}
\eeq
The same contribution $J_{3}$ from (e3) is
\beq
J_{3}=2\beta (\nu -2)A.
\label{cr1-51}
\eeq
Using \reflef{cr1-31}) we compute
\beq
J=J_{1,2}+J_{3}= -\beta\frac{6\alpha}{\pi}m,
\label{cr1-52}
\eeq
which exhibits the same feature of composition-dependence as in (18) of Ref. [\cite{yf1}], but 4 times as large in the magnitude.

In this scenario we have spontaneously broken scale invariance with $\sigma$ as a massless dilaton at the classical level in E frame [\cite{yf5}], while the symmetry is explicitly broken by the effect of a quantum anomaly.  The nonzero coupling as given by \reflef{cr1-52}) eventually results in nonzero dilaton mass, with $\sigma$ now as a pseudo Nambu-Goldstone boson.  This seems to offer yet another theoretical model of a finite-range ``gravitational" force [\cite{yf5}].\footnote{See Ref. [\cite{fsb}], for example, for a review of the phenomenological status of the ``fifth force."}

The same analysis can be applied to the photon field as well.  The diagrams to be included are (a2) (multiplied by $(\nu -2)$) in Fig. 1 and (d2) in Fig. 3.  These are now added together rather than as alternative contributions leading to the same result in BD theory.  We thus have twice the non-trace coupling shown in \reflef{cr1-43}).  The parameters will be constrained by the phenomenological analyses in the same way as in Ref. [\cite{yf1}], but taking the finite force-range into account.

\begin{center}
{\Large\bf References}
\end{center}
\begin{enumerate}
\item\label{yf1}Y. Fujii, Mod. Phys. Lett. {\bf A9}(1994)3685.
\item\label{bd}C. Brans and R.H. Dicke, Phys. Rev. {\bf 124}(1961)925.\item\label{sstr}A. Pais and S.T. Epstein, Rev. Mod. Phys. {\bf 21}(1949)445; J.M. Jauch and F. Rohrlich, {\sl The Theory of Photons and Electrons}, Addison-Wesley, Cambridge, 1954.
\item\label{tran}M.S. Chanowitz and J. Ellis, Phys. Lett. {\bf 40B}(1972)397.
\item\label{tkhs}See also, Y. Fujii and Y. Takahashi, Lecture Notes in Physics 39, {\sl Mathematical Problems in Theoretical Physics,} ed. H. Araki, p.261, Springer-Verlag, Berlin,1975.
\item\label{yf2}Y. Fujii, Brans-Dicke cosmology corrected for a quantum effect due to the scalar-matter coupling, gr-qc/9609044.
\item\label{yf5}Y. Fujii, Int. J. Mod. Phys. {\bf A6}(1991)3505, and papers cited therein.
\item\label{fsb}E. Fischbach and C. Talmadge, Nature, {\bf 356}(1992)207.

\end{enumerate}

\epsfverbosetrue

\begin{figure}[h]
\hspace*{.25cm}
\epsfxsize=14cm
\epsfbox{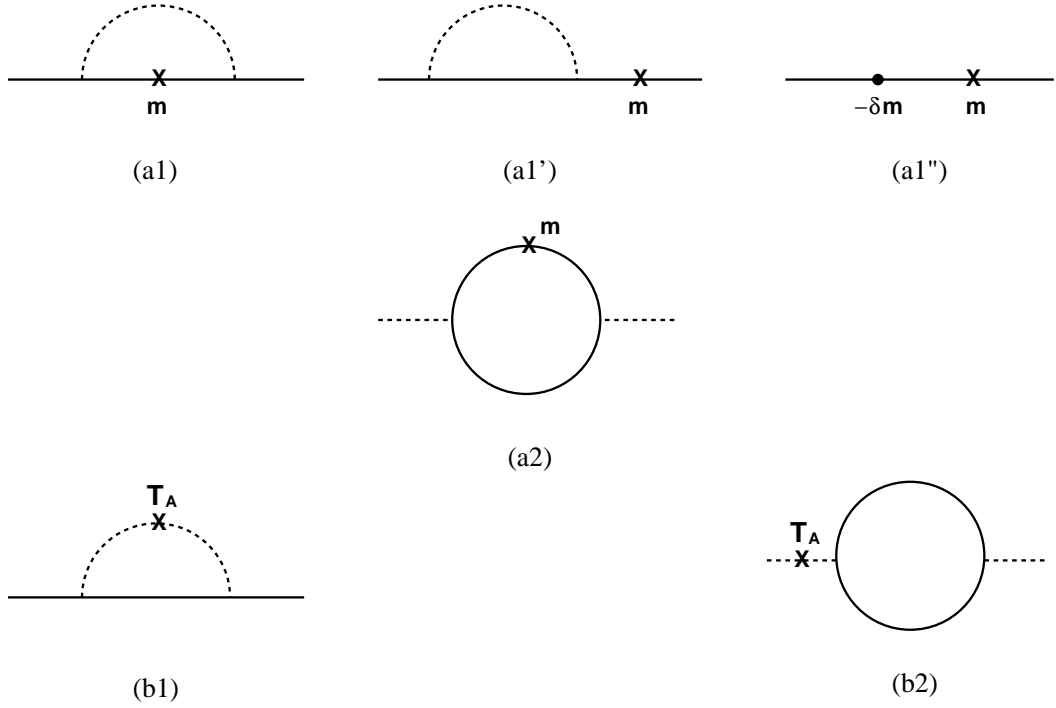}

\caption{Diagrams of 1-loop correction to the $\sigma$-matter coupling.  The $\times$'s in (a's) indicate the mass insertion, first term of (18) ($e^{-\beta\sigma}$ dropped), while those in (b's) are for the second term coming from the photon off 4 dimensions.}
\label{fg1}
\end{figure}

\begin{figure}[h]
\vspace{2cm}
\hspace*{2.5cm}
\epsfxsize=9cm
\epsfbox{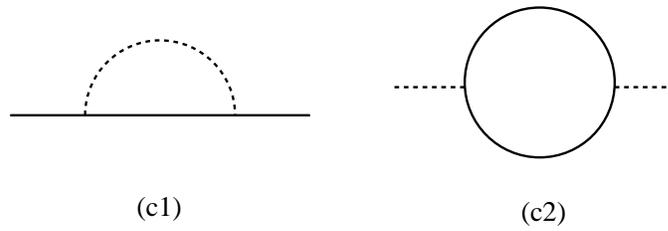}
\caption{Self-energy diagrams.}
\label{fg2}
\end{figure}

\vspace{1cm}
\mbox{}
\mbox{}
\bfg[ht]
\vspace{2cm}
\hspace*{2.5cm}
\epsfxsize=9cm
\epsfbox{ff3.eps}
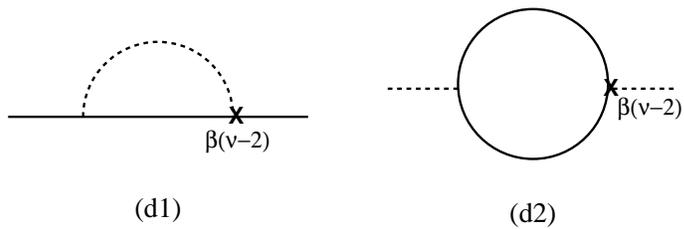
\caption{Diagrams off 4 dimensions.  At each vertex of (c1) and (c2)
now marked by $\times$, $\beta (\nu -2)$ is multiplied.}
\label{fg3}

\efg

\bfg[h]
\epsfxsize=14cm
\epsfbox{ff4.eps}
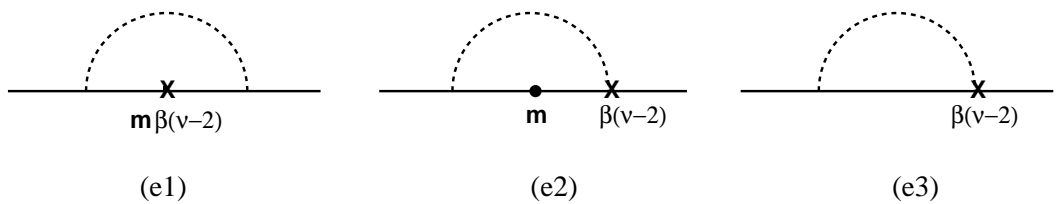
\caption{Diagrams for $\psi$ in the scale-invariant theory. $\beta
(\nu -2)$ is multiplied at $\times$'s. }
\label{fg4}
\efg

\end{document}